%% file: main.tex
\documentclass[a4paper,11pt]{article}
\pdfoutput=1 

\usepackage{jheppub} 

\usepackage[T1]{fontenc}

\title{\boldmath Stable Causality and Microcausality for Drummond–Hathrell Photons}

\author[a]{Madhukar Deb,}
\author[a]{Jay Desai,}
\author[a]{Diptimoy Ghosh}

\affiliation[a]{Department of Physics, Indian Institute of Science, Education and Research, Pune, India}

\emailAdd{madhukar.deb@students.iiserpune.ac.in}
\emailAdd{desai.jay@students.iiserpune.ac.in}
\emailAdd{diptimoy.ghosh@iiserpune.ac.in}

\abstract{Local superluminal photon propagation arises at $\mathcal{O}(\alpha/m_e^2)$ in the Drummond Hathrell (DH) effective action obtained by integrating out the electron in QED coupled to gravity. Whether such superluminality implies a genuine violation of causality in curved spacetime is subtle and remains conceptually nontrivial. In this work we revisit this question using two complementary and largely symmetry-independent diagnostics.

First, we analyse the global causal structure of the effective (optical) metric governing DH photon propagation and identify conditions under which it remains stably causal, thereby excluding the formation of closed causal curves. Second, from a quantum-field-theoretic perspective, we examine microcausality by treating the gravitational background as a fixed Lorentz-breaking field and applying flat-spacetime analyticity bounds to the photon commutator within the geometric-optics regime of the EFT.
  
For two representative examples, a circular photon orbit in Schwarzschild and a linear trajectory in a two-black-hole geometry, we find that, within the regime of validity of the DH effective theory, both diagnostics indicate that the superluminal photon propagation is causally benign. Our results do not constitute a general definition of microcausality in curved spacetime, but provide a controlled and instructive check of causal consistency for EFT superluminality in gravitational backgrounds.}

\begin{document} 
\maketitle
\flushbottom

\section{Introduction}

Relativistic causality places sharp restrictions on the propagation of signals in flat
spacetime: no information can travel outside the Minkowski light cone, and in a quantum
field theory this requirement is encoded in microcausality, namely the vanishing of
operator commutators at spacelike separation. These principles lead to powerful
constraints on effective field theories (EFTs), particularly in settings where Lorentz
invariance and an $S$-matrix description are available (see for example, \cite{Adams_2006, Bellazzini:2014waa, deRham:2017avq,Bellazzini:2017fep,Bellazzini:2020cot,Caron-Huot:2020cmc,Arkani-Hamed:2020blm,Azatov:2021ygj,Bellazzini:2021oaj, Ghosh:2022qqq,Haring:2022sdp,Creminelli:2022onn,Beadle:2025cdx,Desai:2025alt}).

In curved spacetime, however, the relation between signal propagation and causality is
more subtle. Even in a UV-complete and causal theory such as QED coupled to gravity,
integrating out heavy fields generates higher-derivative operators that can deform the
effective causal cone relative to the background metric. A classic example is the
Drummond--Hathrell (DH) effective action \cite{Drummond:1979pp}, in which
${\cal O}(\alpha/m_e^2)$ curvature couplings modify the photon dispersion relation. In
various spacetimes, these corrections permit photon trajectories that are spacelike with
respect to the background metric \cite{Daniels_1994, Daniels_1996, Shore_2003, Goon_2017}.
We refer to this phenomenon as ``superluminality''.

A natural question therefore arises:
\begin{quote}
\emph{Does local superluminal propagation in the DH EFT imply a breakdown of causality?}
\end{quote}
In flat spacetime, the answer would immediately be no because of the equivalence of reference frames, which leads to changing ordering of events. In curved spacetime, the issue is far more intricate. Lorentz invariance is generically broken by the background geometry, reference frames are not equivalent, and familiar flat-space arguments no longer apply. As a result, superluminality does not automatically entail causality violation, a point that has been demonstrated in a variety of curved backgrounds \cite{Geroch:2010da, Papallo_2015, Burrage_2012, Babichev_2008}.

Several works have addressed the causality of DH photons using time-delay arguments \cite{Camanho_2016, de_Rham_2020, Chen_2022}. In situations where an $S$-matrix exists and the geometry enjoys suitable symmetries, DH corrections produce either a small positive time delay across a shockwave \cite{Camanho_2016} or a parametrically negligible negative time delay in the Schwarzschild background \cite{de_Rham_2020, Chen_2022}. While these results are reassuring, time-delay-based reasoning has important limitations. Such arguments rely on asymptotic flatness, the existence of an $S$-matrix, and often strong symmetries such as spherical symmetry, making them difficult to generalise to multi-centred or non-symmetric geometries. Moreover, as emphasised by Penrose \cite{PENROSE19801}, even the operational meaning of ``time delay'' in asymptotically flat spacetimes is subtle, since curved-spacetime null cones can lie outside their flat-space counterparts. Time delay, while useful, is therefore not a universal diagnostic of causality.

These limitations motivate a more geometric and state-independent approach. In this work we adopt two complementary perspectives:
\begin{itemize}
\item[(1)] \textbf{Stable causality:} We analyse the global causal structure of the effective (optical) metric governing DH photon propagation. Stably causal spacetime, which is defined to have no closed causal curves even under small metric perturbations, is equivalent to the existence of a global time function as seen in \cite{Hawking:1973uf, Wald:1984rg}. This approach was previously applied to DH photons in FLRW backgrounds \cite{Shore:2003jx}. Here we extend the analysis to a circular photon orbit in the Schwarzschild geometry and to a straight-line trajectory in a two-centre, extremal Reissner--Nordström background.

\item[(2)] \textbf{Quantum microcausality:} Microcausality is a consequence of the causal structure of a theory rather than of Lorentz invariance itself \cite{Dubovsky_2008}. Using the analyticity framework developed in \cite{Hui_2025, Gavassino_2024}, we examine whether the modified DH dispersion relations are compatible with flat-spacetime microcausality bounds when the gravitational background is treated as a fixed Lorentz-breaking field. We further employ the formalism of \cite{Creminelli:2025rxj} to analyse the structure of Green’s functions in $(\omega,\vec{k})$ space.
\end{itemize}

We apply these diagnostics to two explicit examples exhibiting DH superluminality: (i) a circular photon orbit in Schwarzschild, and (ii) a straight-line trajectory between two extremal Reissner--Nordström black holes. In both cases we find that, within the geometric-optics window of the EFT,
\begin{quote}
\emph{the DH superluminality appears compatible with stable causality and with flat-spacetime microcausality bounds.}
\end{quote}
In the two--black--hole geometry, the stable-causality analysis additionally imposes a parametric condition $M \gg m_e^{-1}$, while the microcausality test remains valid independently of this assumption.

Our aim is deliberately modest. Rather than resolving all causality questions associated with EFT superluminality, we analyse two concrete examples using symmetry-independent and asymptotic-structure-independent diagnostics. This helps clarify why the known DH superluminalities do not lead to causal pathologies in standard gravitational backgrounds, at least within the geometric-optics regime of the effective theory.

The rest of the paper is organised as follows. Section~\ref{sec:DH} reviews the DH action and its superluminal propagation effects. Section~\ref{sec:review} revisits previous time-delay arguments and their limitations. Section~\ref{sec:stable} analyses the stable causality of the optical metric for the Schwarzschild and two black-hole geometries. Sections~\ref{sec:micro} and \ref{micro_omega_k} examine microcausality constraints using the analyticity properties of the Green's functions in $(t, \Vec{k})$ and $(\omega, \Vec{k})$ spaces respectively. We conclude in Section~\ref{Conclusion} with remarks and an outlook.

\subsection*{Convention and Notation} In this paper, we consider $\hbar=c=1$. We also consider that $G=1$. Thus, in the metric, wherever we write $M$, it is $GM$. Also, the metric signature is mostly negative $(+,-,-,-)$. Thus, timelike vectors are those which are such that $g_{\mu\nu}u^\mu u^\nu>0$.\\ 

We denote the background metric as $g_{\mu\nu}$. The general \emph{optical metric} (valid for all photon paths) is denoted by $\mathcal{G}'_{\mu\nu}$ and that for a specific photon path is denoted by $\Tilde{\mathcal{G}}_{\mu\nu}$.

\input{section2}
\input{section3}
\input{section4}
\input{section5}
\input{section6}

\section{Conclusions and Remarks}\label{Conclusion}

In this paper, we revisited the question of whether the superluminal photon propagation
predicted by the Drummond--Hathrell (DH) effective action signals a breakdown of causality
in curved spacetime. Our approach was deliberately modest: rather than attempting to
formulate a fully general criterion, we focused on two complementary and largely
symmetry-independent diagnostics that probe different aspects of causal consistency.

On the classical side, we analysed the global causal structure of the effective optical
metric governing DH photon propagation. For both examples studied—the circular orbit in
Schwarzschild and the straight trajectory in a two--black--hole geometry—we identified
parametric regimes in which the optical metric is stably causal. In these regimes, a
global time function exists, implying the absence of closed causal curves and ruling out
global causal paradoxes. 


On the quantum side, we examined microcausality using the analyticity criteria developed
in \cite{Hui_2025,Creminelli:2025rxj}. Treating the gravitational background as a fixed
Lorentz-breaking field, we tested whether the DH-modified dispersion relations satisfy
certain analyticity and boundedness conditions within the geometric-optics regime of the EFT.
For the trajectories considered, these conditions are satisfied, indicating that the
photon commutator remains compatible with flat-spacetime microcausality bounds. In this
restricted but well-defined sense, the DH corrections do not permit acausal signalling,
despite the local widening of the background light cone.

Taken together, these analyses suggest that DH superluminality is causally benign within
the geometric-optics EFT window,
\begin{equation}
    m_e^{-1}\ll \lambda \ll (R_{\mu\nu\rho\sigma}R^{\mu\nu\rho\sigma})^{-1/4}.
\end{equation}
This does not rule out the possibility of causal pathologies in more extreme regimes or
for other effective field theories. Rather, our results highlight that superluminality in
gravitational EFTs does not \emph{automatically} imply causality violation, and that global
geometric and local quantum diagnostics can jointly provide a sharper picture of when
genuine problems may arise.

\paragraph{Outlook.}
Several directions merit further study. It would be valuable to extend the
stable-causality analysis to more general multi-centre or dynamical geometries, where
the optical metric may exhibit richer causal structure. On the quantum-field-theory
side, one could investigate whether DH-like corrections in other EFTs—for instance those
involving massive higher-spin fields or axion-like couplings—satisfy analogous
analyticity bounds. Finally, it would be interesting to clarify whether a unified
criterion exists that directly links EFT superluminality to geometric properties of the
full UV theory. We hope that the modest results presented here help frame these broader
questions more sharply.

We emphasise, in closing, that our microcausality analysis relies on treating the
gravitational background as a fixed Lorentz-breaking field and applying flat-spacetime
analyticity criteria. While this provides a useful and controlled diagnostic within the
EFT regime, it does not constitute a fully intrinsic definition of microcausality in
curved spacetime. Clarifying such a definition—particularly for theories in which the
effective propagation metric differs from the background geometry—remains an important
open problem.

\appendix
\input{appendix}

\section*{Acknowledgements}
We thank Mohd Ali and Ahmadullah Zahed for insightful comments and discussions. We are grateful to Sunil Mukhi and Suneeta Vardarajan for engaging discussions. DG acknowledges support from the Core Research Grant CRG/2023/001448 of the Anusandhan National Research Foundation (ANRF) of the Gov. of India.

\bibliography{biblio}

\end{document}

%% file: section2.tex
\section{Review of Superluminality in the Drummond-Hathrell Action}
\label{sec:DH}

In their classic analysis \cite{Drummond:1979pp}, Drummond and Hathrell showed that integrating out the electron in QED coupled to gravity generates higher-derivative electromagnetic operators suppressed by $m_e^{-2}$. The resulting effective action, valid well below the electron mass, admits an expansion in powers of $\alpha/m_e^2$,
\begin{align}
    \mathcal{S}=\int d^4x\sqrt{-g}\,\bigg(
    R-\frac{1}{4}F^2
    +a\,R\,F^2
    +b\,R_{\mu\nu}F^\mu_{\ \rho}F^{\nu\rho}
    +c\,R_{\mu\nu\rho\sigma}F^{\mu\nu}F^{\rho\sigma}
    +d\,\nabla_\mu F^{\mu\lambda}\nabla_\nu F^\nu_{\ \lambda}
    \bigg).
\end{align}
The coefficients $(a,b,c,d)$ are all of order $\alpha/m_e^2$, and their explicit values are derived in \cite{Drummond:1979pp, Shore_2003, Goon_2017}. For the purposes of this work, the scaling and sign of these coefficients are more relevant than their precise numerical values.

Varying the action with respect to $A_\mu$ yields the equations of motion
\begin{align}
    \nabla_\mu F^{\mu\nu}
    &-4a\nabla_\mu(RF^{\mu\nu})
    -2b\Big(
        R^\mu_{\ \rho}\nabla_\mu F^{\rho\nu}
        -F^{\rho\mu}\nabla_\mu R^\nu_{\ \rho}
        -R^\nu_{\ \rho}\nabla_\mu F^{\rho\mu}
        +F^{\rho\nu}\nabla_\mu R^\mu_{\ \rho}
    \Big)
    \notag\\
    &\quad
    -4c\nabla_\mu(R^{\mu\nu\rho\sigma}F_{\rho\sigma})
    =0.
\end{align}
The term proportional to $d$ multiplies $\nabla_\mu F^{\mu\nu}$, which itself is ${\cal O}(\alpha/m_e^2)$, so keeping the $d$-term would only contribute at higher order. We therefore consistently drop it in leading-order analyses.

\medskip

In the backgrounds that we consider, the equation of motion is given by:
\begin{align}\label{eq:EOM_RicciFlat}
    \nabla_\mu F^{\mu\nu}-(2b\,R^{\mu}{}_\lambda  \nabla_\mu F^{\lambda\nu}+4c\,R^{\mu\nu\alpha\beta}\nabla_\mu F_{\alpha\beta})=0
\end{align}
Thus, relative to minimally coupled Maxwell theory, the DH terms induce a curvature-dependent correction to the photon propagation.

\subsection*{Geometric optics and the modified light cone}

Superluminal behaviour arises cleanly in the geometric optics approximation. Consider a DH photon with wavelength $\lambda$ propagating in a region of spacetime whose curvature radius is much larger than $\lambda$,
\begin{equation}\label{eq:wavelength_regime}
    m_e^{-1}\ll \lambda \ll (R_{\mu\nu\rho\sigma}R^{\mu\nu\rho\sigma})^{-1/4}.
\end{equation}
This implies a smallness condition
\begin{equation}\label{eq:small_correction}
    \frac{(R_{\mu\nu\rho\sigma}R^{\mu\nu\rho\sigma})^{1/4}}{m_e}\ll 1,
\end{equation}
ensuring that higher-order EFT corrections remain parametrically suppressed.

Under \eqref{eq:wavelength_regime}, the WKB expansion is valid. We write
\begin{equation}
    A_\mu=A\bar{a}_\mu e^{i\Theta},
\end{equation}
with $A\bar{a}_\mu$ being a slowly varying amplitude on the curvature scale and $\Theta$ rapidly varying on the wavelength scale. To leading order,
\begin{equation}
    \partial_\mu \Theta = k_\mu,
\end{equation}
where $k_\mu$ is the photon wavevector, and $\nabla_\mu A_\nu\simeq A \bar a_\nu\partial_\mu e^{i\Theta}$.

Inserting this ansatz into \eqref{eq:EOM_RicciFlat} and multiplying the equation by $\bar{a}_\nu$ gives us:
\begin{equation}\label{eq:generalEOM}
    k^2-2b\,R^{\mu\nu}k_\mu k_\nu+8c\,R^{\mu\alpha\nu\beta}k_\mu k_\nu \bar{a}_\alpha\bar{a}_\beta=0
\end{equation}
working in Lorenz gauge $k_\mu \bar a^\mu=0$, with normalization $\bar a^\mu \bar a_\mu=-1$ (so $\bar a_\mu$ is spacelike). Here, we have assumed that the background $A_\mu$ field is absent. If we assume that the background field $\bar{A}_\mu$ is non-zero, then the above equation is modified by:
\begin{align}
    k^2-2b\,R^{\mu\nu}k_\mu k_\nu+8c\,R^{\mu\alpha\nu\beta}k_\mu k_\nu \bar{a}_\alpha\bar{a}_\beta+4c\,R^{\mu\nu\alpha\beta}\nabla_\mu \bar{F}_{\alpha\beta}\bar{a}_\nu=0
\end{align}

The curvature-dependent term suggests defining an \emph{effective} or \emph{optical} inverse metric $\tilde{g}^{\mu\nu}$ such that
\begin{equation}\label{eq:optical_metric}
    \tilde{g}^{\mu\nu}=g^{\mu\nu}-2b\,R^{\mu\nu}+8c\,R^{\mu\rho\nu\sigma}\bar a_\rho \bar a_\sigma
\end{equation}
for which the propagation condition takes the simple form
\begin{equation}
    \tilde{g}^{\mu\nu}k_\mu k_\nu=4c\,R^{\mu\nu\alpha\beta}\nabla_\mu \bar{F}_{\alpha\beta}\bar{a}_\nu
\end{equation}
Equivalently, the effective metric $\tilde{\mathcal G}_{\mu\nu}$ satisfies as seen in \cite{Shore_2003, Jing_2016}
\begin{equation}\label{eq:opt_null}
    \tilde{\mathcal G}_{\mu\nu}u^\mu u^\nu=4c\,R^{\mu\nu\alpha\beta}\nabla_\mu \bar{F}_{\alpha\beta}\bar{a}_\nu
\end{equation}
with $\tilde{\mathcal G}_{\mu\alpha}\tilde{g}^{\alpha\nu}=\delta^\nu_{\ \mu}$. To leading order,
\begin{equation}\label{eq:inverse_metric}
    \tilde{\mathcal G}_{\mu\nu}=g_{\mu\nu}+2b\,R_{\mu\nu}-8c\,R^{\rho}{}_{\mu}{}^{\sigma}{}_{\nu}\,\bar a_\rho \bar a_\sigma
\end{equation}
with corrections controlled by \eqref{eq:small_correction}. In the spacetimes we will consider, the right hand side of \eqref{eq:opt_null} is 0. Hence, the DH terms slightly deform the causal cone of $g_{\mu\nu}$, while the photon remains null with respect to $\tilde{\mathcal G}_{\mu\nu}$. \\

This effective metric is obtained for a particular path $\gamma$ ($\tilde{\mathcal{G}}^{\gamma}_{\mu\nu}$). For a general path, the effective metric ($\mathcal{G}'_{\mu\nu}$) is defined by
\begin{equation}
    \mathcal{G}'_{\mu\nu}|_\gamma=\tilde{\mathcal{G}}^{\gamma}_{\mu\nu}
\end{equation}
for all paths $\gamma$.
\medskip
 In pure Maxwell theory the causal structure is governed by $g_{\mu\nu}$, and $g^{\mu\nu}k_\mu k_\nu=0$ encodes both null propagation and microcausality. The DH modification does \emph{not} alter the fact that the photon moves on a null trajectory relative to some metric; it merely shifts the null cone from that of $g_{\mu\nu}$ to that of $\tilde{\mathcal G}_{\mu\nu}$. Thus, ``superluminality'' means that a photon trajectory may be spacelike with respect to $g_{\mu\nu}$ but remains lightlike with respect to $\tilde{\mathcal G}_{\mu\nu}$.

\medskip
The geometric optics window \eqref{eq:wavelength_regime} is essential. For $\lambda\lesssim m_e^{-1}$ the DH effective action breaks down and one must revert to the UV theory (QED with gravity), while for $\lambda\gtrsim (R_{\mu\nu\rho\sigma}R^{\mu\nu\rho\sigma})^{-1/4}$ geometric optics fails and wave effects become important. The conclusions about superluminality and optical metrics only apply inside the EFT window.

\subsection*{Example I: circular photon orbit in Schwarzschild}

Consider an equatorial circular photon orbit around a Schwarzschild black hole of mass $M$,
\begin{equation}
    ds^2=\Big(1-\frac{2M}{r}\Big)dt^2
    -\Big(1-\frac{2M}{r}\Big)^{-1}dr^2
    -r^2 d\theta^2
    -r^2\sin^2\theta\, d\phi^2.
\end{equation}
Along a circular null trajectory $u^r=u^\theta=0$, with $u^t,u^\phi\neq 0$. In this background the Riemann tensor components relevant to our polarization choice satisfy
\begin{equation}
    R^{\mu\rho}{}_{\nu\sigma}
    =\begin{cases}
        R^{\mu\rho}{}_{\mu\rho} &\mu=\nu,\ \rho=\sigma,\\
        0 &\text{otherwise}.
    \end{cases}
\end{equation}
Considering the propagation of the photon in the $\phi$ direction in the plane $\theta=\pi/2$, with a radial polarization, we see that:
\begin{align}
    \bigg(1-\frac{2M}{r}\bigg)(1+8cR^{t r}_{ \ \ tr})\bigg(\frac{dt}{d\tau}\bigg)^2-r^2(1+8cR^{\phi r}_{ \ \ \phi r})\bigg(\frac{d\phi}{d\tau}\bigg)^2=0
\end{align}
Now, we can rewrite the above as:
\begin{align}
     \bigg(1-\frac{2M}{r}\bigg)\bigg(\frac{dt}{d\tau}\bigg)^2-r^2\bigg(\frac{d\phi}{d\tau}\bigg)^2&=8c\bigg(r^2R^{\phi r}_{ \ \ \phi r}\bigg(\frac{d\phi}{d\tau}\bigg)^2\notag\\
     &-\bigg(1-\frac{2M}{r}\bigg)R^{t r}_{ \ \ tr}\bigg(\frac{dt}{d\tau}\bigg)^2\bigg)
\end{align}
Substituting the Riemann tensor values
\begin{equation}
    R^{\phi r}{}_{\phi r}=\frac{M}{r^3},\qquad
    R^{t r}{}_{tr}=-\frac{2M}{r^3},
\end{equation}
and with $c<0$ we obtain
\begin{equation}
    g_{\mu\nu}u^\mu u^\nu<0,
\end{equation}
so the trajectory is spacelike relative to the background metric. Locally, the photon appears ``superluminal'' with respect to $g_{\mu\nu}$.

\subsection*{Example II: straight line between two extremal black holes}

Next consider a multicenter Reissner-Nordström configuration with two extremal black holes at $(0,0,\pm L)$. The spacetime metric is \cite{carroll2004spacetime}
\begin{equation}\label{eq:R-N}
    ds^2=U^{-2}(\vec{x})\,dt^2-U^2(\vec{x})\,d\vec{x}^2,\qquad
    U(\vec{x})=1+\frac{M}{\Delta X_+}+\frac{M}{\Delta X_-},
\end{equation}
with $\Delta X_\pm=\sqrt{(L\mp z)^2+x^2+y^2}$. In this background, equation \eqref{eq:EOM_RicciFlat} applies. In this case, the EM background $\bar F_{\mu\nu}$ will be non-zero. From \cite{carroll2004spacetime}, we find that
\begin{equation}\label{eq:EM_Background}
    \bar A_0 = U^{-1}-1,~~~ \bar A_i =0
\end{equation}
With this background, and considering polarization along a single ($z$) direction, the RHS of \eqref{eq:opt_null} vanishes due to the fact that $R^{tijk}=0$. 
Consider propagation along $x$ with polarization along $z$. To leading order, one finds
\begin{align}
    U^{-2}(1+8c\, R^{tz}{}_{tz}+2b\,R^{t}{}_t)\Big(\frac{dt}{d\tau}\Big)^2
    -U^{2}(1+8c\, R^{xz}{}_{xz}+2b\,R^{x}{}_x)\Big(\frac{dx}{d\tau}\Big)^2=0.
\end{align}
Matching to the background null condition $U^2 dx/d\tau = U^{-2} dt/d\tau$ yields
\begin{equation}
    U^{-2}\Big(\frac{dt}{d\tau}\Big)^2
    -U^{2}\Big(\frac{dx}{d\tau}\Big)^2
    =(8c\,(R^{xz}{}_{xz}-R^{tz}{}_{tz})-2b\,(R^{t}{}_t-R^{x}{}_x))\,
    U^{-2}\Big(\frac{dt}{d\tau}\Big)^2.
\end{equation}
But since $R^{t}{}_t=R^{x}{}_x$ in the X-Y plane, the second term on the right-hand side is 0. 

Using the values of the Riemann tensor given in \cite{Goon_2017}, we get
\begin{equation}
    R^{xz}{}_{xz}-R^{tz}{}_{tz}
    =\frac{6U^{-2}L^2M}{(L^2+x^2)^{5/2}}>0,
\end{equation}
and because $c<0$, one again finds
\begin{equation}
    g_{\mu\nu}u^\mu u^\nu<0,
\end{equation}
so the trajectory is spacelike relative to $g_{\mu\nu}$. Superluminality is therefore also present in this multicenter geometry.

\medskip
 The appearance of local superluminal propagation is familiar from flat spacetime as an immediate signal of acausality: propagation outside the Minkowski light cone conflicts with microcausality and allows signal transmission outside causal order. The DH setup therefore raises an important conceptual question: if a consistent UV theory such as QED coupled to gravity generates small polarization-dependent superluminal effects at low energies, does this necessarily endanger causality in a curved background? In gravitational settings the notion of causal structure is richer, and the ``background'' light cone need not universally control signal propagation. The central issue investigated in later sections is whether such local superluminality can accumulate into a global causal pathology, or whether it remains entirely benign once the full spacetime structure is taken into account. We now turn to causality-based diagnostics aimed at addressing this question.

%% file: section3.tex
\section{Superluminality and Causality: A Review of Time-Delay Arguments}
\label{sec:review}

As shown in the previous section, the Drummond-Hathrell (DH) corrections cause photon trajectories to deviate slightly from the null cone of the background metric $g_{\mu\nu}$. In flat spacetime even an infinitesimal superluminal deviation would immediately signal a breakdown of relativistic causality, since it allows information to propagate outside the Minkowski light cone. In curved spacetime, however, the conceptual landscape is richer: different effective metrics may govern propagation, the background lacks global Lorentz symmetry, and the meaning of ``delay'', ``advance'', or ``signal speed’’ depends sensitively on the geometry. This has led to a substantial literature exploring whether DH superluminality is truly dangerous or merely an innocuous EFT artifact.

One widely studied diagnostic is the \emph{time delay}. Intuitively, one asks whether a photon arrives earlier or later than it would have in some reference geometry (often flat spacetime or minimally coupled Maxwell theory).\footnote{Here ``time delay'' refers to a quantity extracted from a coordinate that plays the role of time in an appropriate asymptotic region. The interpretation is therefore tied to the asymptotic structure of the spacetime.} Positive time delay is traditionally associated with causal propagation, while negative time delay (“time advance”) is often regarded as a red flag for causality violation. In this section we review two influential lines of reasoning that employ time delay to argue that DH superluminality does not jeopardize causality, and then explain why time delay is not reliable as a general criterion. This motivates the use of more robust causal diagnostics in the rest of the paper.

\subsection*{Shockwave Backgrounds and Local Shifts in Null Coordinates}

In \cite{Camanho_2016}, the authors study a DH photon in a gravitational shockwave background in $D$ dimensional space described by the Aichelburg-Sexl metric \cite{Aichelburg:1970dh},
\begin{align}
    ds^2 = du\,dv - h(u,x_i)\,du^2 - \sum_{i=1}^{D-2} dx_i^2,
\end{align}
with $v$ a null Killing direction and $r=\sqrt{x_ix_i}$. By arranging two shockwaves localized at $r=\pm b$ and $u=0$, one can track a photon traveling along the $u$ direction at $r=0$. The relevant observable is the discontinuity
\begin{equation}
    \Delta v = v(u>0) - v(u<0),
\end{equation}
interpreted as the time delay across the shock.

The result is
\begin{align}
    \Delta v=K\bigg(1+c\frac{(D-2)(D-4)}{b^2}
    \bigg(\frac{\vec{a}\!\cdot\!\vec{n}}{a^2}-\frac{1}{D-2}\bigg)\bigg),
\end{align}
with $K>0$, $c\sim \alpha/m_e^2$, $\vec{a}$ the polarization, and $\vec{n}=\vec{b}/b$. The DH correction is suppressed provided the geometric optics condition
\begin{equation}
    m_e^{-1} \ll \lambda \ll b,
\end{equation}
holds, ensuring $(m_e b)^{-1}\ll 1$. Thus the DH term is always a small correction, and $\Delta v$ remains strictly positive in the EFT regime. So we see that no time advance occurs in this case.

\subsection*{Eisenbud-Wigner Scattering Time Delay}

A conceptually different notion of time delay arises in scattering theory and is well defined only in asymptotically flat spacetimes. In \cite{de_Rham_2020,Chen_2022}, the authors consider the Eisenbud-Wigner time delay:
\begin{align}
    \Delta T_\ell = 2\,\frac{\partial\delta_\ell(\omega)}{\partial \omega},
\end{align}
where $\delta_\ell(\omega)$ is the phase shift in the $\ell$-th partial wave.

For the DH photon,
\begin{align}
    \Delta T_\ell = \Delta T_\ell^g + \Delta T_\ell^{\text{EFT}},
\end{align}
where $\Delta T_\ell^g>0$ is the usual time delay for Schwarzschild. Causality requires
\begin{equation}\label{eq:EFT_small_delay_condition}
    |\Delta T_\ell^{\text{EFT}}|\ll \omega^{-1},
\end{equation}
so that the EFT correction cannot overwhelm the positive gravitational delay. The DH contribution scales as
\begin{equation}
    |\Delta T_\ell^{\text{EFT}}|\sim \frac{M}{b^2 m_e^2},
\end{equation}
and in the geometric optics regime ($b\gg M$, $\omega b\gg 1$),
\begin{equation}
    |\Delta T_\ell^{\text{EFT}}|
    \ll \sqrt{\frac{M}{b}}\,\omega^{-1}
    < \omega^{-1}.
\end{equation}
Thus the DH photon again exhibits no time advance in this setting.

\subsection*{Why Time Delay Is Not a Universal Causality Diagnostic}

While the above examples are reassuring, time delay suffers from important conceptual and geometric limitations.

\paragraph{(1) Issues with asymptotically flat spacetimes:} Both the papers \cite{de_Rham_2020} and \cite{Camanho_2016} use time delay in asymptotically flat spacetimes to argue that causality holds. While calculating the time delay, both papers use the $S$-matrix formalism either explicitly (in the case of \cite{de_Rham_2020}) or implicitly by using scattering amplitudes to calculate the time delay. But, in asymptotically flat spacetimes, we have to be very careful how we define $S$-matrices, because, as pointed out in \cite{PENROSE19801}, we cannot define an $S$-matrix without violating the timelike vectors going outside the flat spacetime light cone. So the condition of a positive time delay is too restrictive and may lead to us calling causal spacetimes acausal due to this time delay condition. 

Also, as shown by Gao and Wald in \cite{Gao_2000}, under the null energy condition, null geodesics in any asymptotically AdS spacetimes are always delayed relative to null geodesics in pure AdS. This is a rigorous time delay theorem that works only for asymptotically AdS spacetimes. There is no such equivalent theorem for all asymptotically flat spacetimes. Therefore, time delay may not be the most appropriate quantity to understand causality of asymptotically flat spacetimes.

\paragraph{(2) Spacetime Symmetries:}The argument made in \cite{de_Rham_2020} requires that the spacetime has spherical symmetry so we can define the scattering phase shift for each partial wave $\ell$. This may not work for even a general asymptotically flat spacetime, as seen in the case of the two extremal Reissner-Nordstrom black hole case, where spherical symmetry is broken and thus, we cannot use the concept of scattering time delay using the scattering phase shift.

\medskip
These reasons motivate the approach taken in the remainder of this paper, where we try to assess the causality in a symmetry and background-independent manner. The two particular methods that we will discuss in the next two sections are given below

\begin{itemize}
    \item \textbf{Stable causality} of the optical metric, a global geometric condition ensuring the absence of closed causal curves even under metric perturbations.
    \item \textbf{Microcausality} of the photon field, a quantum condition requiring vanishing of commutators at spacelike separation even in Lorentz-breaking curved backgrounds.
\end{itemize}

These provide a more robust and general framework for analyzing causality in the Drummond-Hathrell effective theory. 

%% file: section4.tex
\section{Stable Causality of the Optical Metric}
\label{sec:stable}

In the previous section, we reviewed how the Drummond-Hathrell (DH) corrections lead to local superluminal propagation in two explicit geometries: a circular photon orbit in Schwarzschild, and a linear trajectory in the field of two extremal Reissner–Nordström black holes. Local superluminality by itself does not immediately diagnose a breakdown of causality in general curved spacetimes due to the arguments made for flat spacetime not translating to curved spacetime. In curved spacetime, the appropriate criterion for causal well-definedness is inherently global.

A spacetime can be defined as \emph{causal} if it contains no CCCs. However, checking this directly is highly nontrivial. Instead, following \cite{Hawking:1973uf,Wald:1984rg}, one typically employs the stronger condition of \emph{stable causality}. A spacetime $(\mathcal{M}, g_{\mu\nu})$ is stably causal if it does not admit CCCs
even after small metric perturbations. Following the proof from \cite{Hawking:1973uf}, a spacetime is stably causal iff there exists a smooth function $f$ whose gradient is everywhere timelike. Operationally, $f$ serves as a global time function: if such an $f$ exists, then the spacetime does not admit CCCs and remains non-pathological under arbitrarily small perturbations of the metric. This analysis was also done for the FLRW metric in \cite{Shore:2003jx}, where Shore found that a DH photon propagating in the FLRW metric is consistent with stable causality.

In our context, the relevant geometry is not the background metric $g_{\mu\nu}$ but the \emph{effective} (or “optical”) metric $\mathcal{G}'_{\mu\nu}$ governing DH photon propagation. For each causal curve of the photon, $\mathcal G'_{\mu\nu}$ reduces to the optical metric $\tilde{\mathcal{G}}_{\mu\nu}$ defined in \eqref{eq:inverse_metric}. Throughout this section, the manifold $\mathcal{M}$ denotes the region exterior to black-hole horizons, where the EFT and geometric optics approximations remain valid.

Our aim is modest: to identify regimes in which $\mathcal G'_{\mu\nu}$ is stably causal. This is sufficient to show that, in those regimes, the DH photon—although superluminal relative to $g_{\mu\nu}$—does not generate closed causal curves with respect to the physical propagation metric.

\subsection{Schwarzschild Geometry}

For Schwarzschild, the effective metric for all DH photons propagating in this spacetime can be brought to the following form as shown by \cite{Drummond:1979pp}:
\begin{align}\label{26}
    ds^2 = \left( 1 - \beta \frac{\alpha}{m_e^2}\frac{M}{r^3} \right)
    \big( F(r)\, dt^2 - F^{-1}(r) dr^2 \big)
    - r^2 d\Omega^2 ,
\end{align}
where $F(r)=1-2M/r$ and $\beta\sim\mathcal{O}(1)$ ($\beta$ is a metric parameter which depends on polarization and direction of propagation).

To test stable causality, we choose the natural candidate
\begin{align}
    f(x^\mu) = t .
\end{align}
The gradient $\nabla_\mu f$ is purely timelike if $g'^{tt}>0$ everywhere ($g'^{\mu\nu}$ is the inverse of $\mathcal G'_{\mu\nu}$). From \eqref{26},
\begin{align}
    g'^{tt} = 1 + \beta \frac{\alpha}{m_e^2}\frac{M}{r^3}.
\end{align}
Using the EFT inequality \eqref{eq:wavelength_regime} and the fact that $R_{\mu\nu\rho\sigma}\propto M/r^3$, we find $g'^{tt}>0$ for all $r>2M$ and for all polarizations.

Thus $t$ is a global time function on $(\mathcal{M}, \mathcal{G}'_{\mu\nu})$, establishing stable causality in the EFT regime. In particular:
\begin{itemize}
    \item the optical metric does not admit closed causal curves,
    \item and the superluminal DH photon remains globally causal even though its null cone is slightly wider than that of $g_{\mu\nu}$.
\end{itemize}

This result does not rule out all possible causal issues in more exotic regions of the parameter space, but it confirms that within the geometric-optics EFT window, no causal pathology arises in Schwarzschild due to DH corrections.

\subsection{Two Black–Hole Geometry}

For the two–center extremal Reissner–Nordström configuration, the explicit effective metric $\mathcal{G}'_{\mu\nu}$ is not known in closed form. Nevertheless, its causal properties can be inferred directly from the structure of the DH correction.

From \eqref{eq:optical_metric}, along any causal photon trajectory $\gamma$,
\begin{align}
    g'^{\mu\nu}|_\gamma = g^{\mu\nu}-2b\,R^{\mu\nu} + 8c R^{\mu\alpha\nu\beta} \bar a_\alpha \bar a_\beta .
\end{align}
Thus,
\begin{align}
    g'^{tt}|_\gamma = g^{tt}-2b\,R^{tt} + 8c R^{t\alpha t\beta} \bar a_\alpha \bar a_\beta .
\end{align}

Near either black hole, the curvature satisfies $R^{\mu\alpha\nu\beta} \bar a_\alpha \bar a_\beta \sim 1/M^2$, the maximum scale in the geometry. Suppose we again take $f(x)=t$. Then
\begin{align}
    g'^{tt} \gtrsim 
    U^2(\vec{x}) \left( 1 + \gamma\, U^{-2}(\vec{x})\, \frac{1}{M^2} \right),
\end{align}
with $U(\vec{x})>1$ in the exterior region and $\gamma\sim\mathcal{O}(\alpha/m_e^2)$ ($\gamma$ is a parameter depending on the polarization and direction of propagation).

Demanding stable causality, we find the condition $\gamma/M^2 \sim \alpha/(m_e^2 M^2) \ll 1$ which is equivalent to
\begin{align}
    m_e M \gg 1 ,
\end{align}
meaning that the BH mass is well above the mass scale $M_{Pl}^2/m_e\sim 10^{-13}M_\odot$. In this regime,
\begin{align}
    \left| \gamma\, U^{-2}(\vec{x})\,\frac{1}{M^2} \right| \ll 1 ,
\end{align}
independent of its sign.

Hence $g'^{tt}>0$ everywhere outside the horizons, and $t$ again defines a global time function. We therefore conclude:

\begin{quote}
\emph{In the two–black–hole geometry, whenever $M \gg m_e^{-1}$, 
the optical metric governing DH photon propagation is stably causal. Local superluminality does not lead to closed causal curves.}
\end{quote}

We emphasize that this is not a proof of causality for all parameter values; rather, it identifies a clean parametric window where stable causality is ensured.

%% file: section5.tex
\section{Microcausality Bounds in the EFT Regime}
\label{sec:micro}

Stable causality of the optical metric provides a classical diagnostic of whether
Drummond--Hathrell (DH) superluminality can endanger the global causal structure of
a spacetime. A complementary question is whether DH photon propagation is compatible
with \emph{microcausality}, the quantum condition that local operators commute at
spacelike separation. Unlike stable causality, microcausality probes the structure
of local observables and is sensitive to the analytic properties of Green’s functions.

In this section we apply the formalism developed in
\cite{Hui_2025, Gavassino_2024} to obtain a modest but informative check of quantum
causal consistency within the DH effective field theory. Our analysis is restricted
to the geometric-optics regime of the EFT and should be understood as a controlled
diagnostic rather than a fully covariant treatment of quantum fields in curved
spacetime.

\subsection{Microcausality and the Paley--Wiener Theorem}

In flat spacetime, microcausality requires that for any two bosonic operators
$\mathcal{O}_1(x)$ and $\mathcal{O}_2(y)$,
\begin{align}
    [\mathcal{O}_1(x),\mathcal{O}_2(y)] = 0
    \qquad \text{if } (x-y)^2 < 0 .
\end{align}
For a field operator $\hat{\phi}(x)$, this condition implies that the commutator
\begin{align}
    G_c(t_1,\vec{x}_1;t_2,\vec{x}_2)
    := [\,\hat{\phi}(x_1), \hat{\phi}(x_2)\,]
\end{align}
vanishes outside the flat-spacetime light cone.

Using the Paley--Wiener theorem \cite{strichartz2003guide}, the above spacetime
statement can be reformulated as a condition on the analyticity and boundedness of
the spatial Fourier transform $\tilde{G}_c(\Delta t,\vec{k})$ at fixed $\Delta t$.
As shown in \cite{Hui_2025}, flat-spacetime microcausality holds if and only if
$\tilde{G}_c(\Delta t,\vec{k})$ admits an analytic continuation to all
$\vec{k}\in\mathbb{C}^3$ and satisfies
\begin{align}\label{PWBound}
    |\tilde{G}_c(\Delta t,\vec{k})|
    \le C(|\vec{k}|)\, e^{\mathrm{Im}(|\vec{k}|)\,\Delta t}
    \qquad \forall \vec{k}\in\mathbb{C}^3 ,
\end{align}
for some polynomial $C$. Importantly, this is an operator statement: it must hold for
\emph{all} states of the theory, including Lorentz-breaking states.

In what follows, we treat the gravitational background as a fixed Lorentz-breaking
state associated with a non-trivial metric configuration. In this sense, the metric
is viewed as a classical spin--2 field living on flat spacetime, analogous to a
background medium, and we examine whether the modified DH dispersion relations
respect the analyticity bound \eqref{PWBound} within the geometric-optics window of
the EFT.

Before proceeding, it is important to clarify the scope of this analysis. In a fully
curved spacetime, there is no unique or universally accepted definition of
microcausality for effective field theories whose characteristic cones differ from
that of the background metric. In particular, for DH photons the relevant
propagation cones are governed by the optical metric rather than $g_{\mu\nu}$.
Establishing a fully covariant notion of microcausality in such settings remains a
subtle and open problem.

Here we therefore adopt a more limited but well-defined perspective. We apply
flat-spacetime microcausality criteria—formulated as analyticity and boundedness
conditions on momentum-space Green’s functions—to the photon field propagating on a
fixed Lorentz-breaking gravitational background. This approach follows the framework
of \cite{Hui_2025, Gavassino_2024} and allows us to test whether the DH-modified
dispersion relations are compatible with standard microcausality bounds within the
regime of validity of the EFT.

Our goal is not to propose a general definition of microcausality in curved
spacetime, but rather to provide a controlled and instructive check: whether
superluminal DH photon propagation, viewed as a Lorentz-breaking effective
excitation on flat spacetime, violates flat-space microcausality bounds in the
geometric-optics regime.

\subsection{Circular Orbit in the Schwarzschild Geometry}

Using the geometric-optics ansatz, in appendix \ref{appendix:A} we derive that the form of the field $A_\mu$ must be as follows:
\begin{align}
    A_\mu = \int \bar{a}_\mu^k \, e^{i(\omega t - k r \phi)} \, dk,
\end{align}
and substituting into the DH equation of motion \eqref{eq:EOM_RicciFlat}, we obtain the modified dispersion relation
\begin{align}
    \omega^2 F^{-1}(1 - 8c R^{tr}{}_{tr})
    - k^2 (1 - 8c R^{\phi r}{}_{\phi r}) = 0.
\end{align}
Solving for $\omega$ yields
\begin{align}\label{eq:Schwarzschild_omega}
    \omega = k \,
    \frac{1 - 4c R^{\phi r}{}_{\phi r}}
         {1 - 4c R^{tr}{}_{tr}}
    F^{1/2}(r),
\end{align}
with $F(r)=1-2M/r$. The commutator in momentum space (Eq (\ref{SchwarzchildG})) behaves as
\begin{align}
    ({\tilde{G}_{\mu\nu}})_c(t,k) \sim i \frac{\sin(\omega t)}{\omega}.
\end{align}
Microcausality Eq (\ref{PWBound}) requires $\mathrm{Im}(\omega)\le \mathrm{Im}(k)$, which gives the condition
\begin{align}\label{MicroIneqSch}
    \frac{1 - 4c R^{\phi r}{}_{\phi r}}
         {1 - 4c R^{tr}{}_{tr}}
    F^{1/2}(r)
    \le 1.
\end{align}
In the Schwarzschild background one finds
\begin{align*}
    1 \le 
    \frac{1 - 4c R^{\phi r}{}_{\phi r}}
         {1 - 4c R^{tr}{}_{tr}}
    \le \bigg(1 - \frac{2M}{r}\bigg)^{-1/2},
\end{align*}
while $F^{1/2}(r) \le 1$. 

Thus the inequality \eqref{MicroIneqSch} is satisfied throughout the
geometric-optics EFT regime. Although DH photons are locally superluminal
relative to the background metric $g_{\mu\nu}$, the modified dispersion
relation does not violate the Paley--Wiener analyticity bound
\eqref{PWBound}. In this sense, when treated as a Lorentz-breaking effective
excitation on a fixed background, the photon commutator remains compatible
with flat-spacetime microcausality within the regime of validity of the EFT.

\subsection{Linear Trajectory in the Two Black–Hole Geometry}

For the two-center extremal Reissner-Nordström configuration,  the geometric-optics ansatz takes the form (shown in appendix \ref{appendix:A})
\begin{align}
    A_\mu = \int \bar{a}^k_\mu \, e^{i(\omega t - k x^*)} \, dk,
\end{align}
with the effective $X$ coordinate
\begin{align}\label{eq:x*}
    x^* = \int^x dx'\,
    \frac{1 - 4c\, R^{tz}{}_{tz}-b\, R^{t}{}_{t}}{1 - 4c\, R^{xz}{}_{xz}-b\, R^{x}{}_{x}}U^2(x')
\end{align}
where $U(x)=U(({x},0,0))$ is defined in \eqref{eq:R-N}. In this case the dispersion relation simplifies to
\begin{align}
    k = \omega .
\end{align}
Consequently,
\begin{align}
    ({\tilde{G}_{\mu\nu}})_c(t,k) \sim i \frac{\sin(kt)}{k},
\end{align}
which is easily bounded by a function satisfying \eqref{PWBound}. The commutator has compact support within the flat-space light cone, and the analyticity requirements of microcausality are met inside the EFT window.

\subsection*{Remarks on the Regime of Validity}

The above conclusions apply only within the geometric-optics EFT regime,
\begin{align}
    l^{-1} \ll |\vec{k}| \ll m_e,
\end{align}
where $l^{-1}\sim (R_{\mu\nu\rho\sigma}R^{\mu\nu\rho\sigma})^{1/4}$ is the curvature scale.
For $|\vec{k}|\lesssim l^{-1}$ the geometric-optics approximation fails, while for
$|\vec{k}|\gtrsim m_e$ the DH EFT is no longer valid and the UV completion (QED coupled
to gravity) must be used. Since the UV theory is known to be causal
\cite{Hollowood_2016, Hollowood_2007}, one does not expect any fundamental causal
inconsistency to arise even outside the EFT regime, although the effective description
employed here is no longer applicable.

Our results should therefore be viewed as a limited but constructive check:
\begin{quote}
\emph{Within the window of validity of the Drummond--Hathrell effective theory, the
modified dispersion relation appears compatible with the microcausality bounds
derived from the Paley--Wiener theorem, at least for the explicit trajectories
examined here.}
\end{quote}
This does not constitute a general proof of microcausality for DH photons in arbitrary
backgrounds, but it provides additional evidence that local superluminality in the DH
EFT need not entail a causal pathology.

%% file: section6.tex
\section{Microcausality from the Green's function in \text{$(\omega,\vec{k})$} space}\label{micro_omega_k}
Now, that we have discussed the causality condition derived from the mixed space $(t,\vec{k})$ Green's function for superluminality that is derived from the DH action, we can use the Fourier transformed Green's function in $(\omega,\vec{k})$ space to derive the causality conditions. To do this, we will use the formalism developed in the recently published paper \cite{Creminelli:2025rxj} to see if our Green's function is consistent with causality.

Considering $G_\text{R}(\omega,\vec{k})$ is the retarded Green's function in the $(\omega,\vec{k})$ space, then according to \cite{Creminelli:2025rxj}, $\mathrm{Im}G_\text{R}(\omega,\vec{k})$ will be microcausal if  
\begin{align}\label{condition}
    \int d\zeta \ \mathrm{Im}G_\text{R}(\zeta,{k}+\xi\zeta)=0
\end{align}
which is shown to be true for the Lorentz invariant case. For the Lorentz invariant case, we see that $\mathrm{Im}G_\text{R}(\omega,\vec{k})$ has the following form:
\begin{align}\label{imG}
    \mathrm{Im}G_\text{R}(\omega,\vec{k})=\mathrm{Sign}(\omega)\theta(\omega^2-k^2)\rho(\omega^2-k^2)
\end{align}
where $\rho(\omega^2-k^2)$ is the spectral density and $\theta(\omega^2-k^2)$ is the Heaviside step function.

In \cite{Creminelli:2025rxj}, the authors show that if the Green's function is given by \eqref{imG}, then it follows equation \eqref{condition} and thus, is consistent with causality. They also show that if the spectral density is of the form $\rho(\omega^2-c_s^2k^2)$ (spectral density for fields in a Lorentz broken background), then this Green's function is also consistent with causality as long as $c_s^2\leq1$.

The spectral density in the EFT regime for the DH action is of the form $\rho(\omega^2-c_s^2k^2)$ as the background breaks Lorentz invariance. We will show that $c_s^2\leq1$ for each of the cases, which will imply that the superluminality seen in the DH action is consistent with causality. For the single DH photon in curved backgrounds, the spectral density is of the form of a delta function. 

\subsection*{Circular Path in Schwarzschild}
In this section, we will find the value of $c_s^2$ for a superluminal DH photon propagating in a circular path in the Schwarzschild metric. The value of $c_s$ is given in equation \eqref{eq:Schwarzschild_omega}. Thus, we see that:
\begin{align}
    c_s^2=\frac{1 - 8c R^{\phi r}{}_{\phi r}}
         {1 - 8c R^{tr}{}_{tr}}
    F(r)
\end{align}
Thus, the spcetral density $\rho$ is given by:
\begin{align}
    \rho(\omega^2-c_s^2k^2)=2\pi\delta({\omega^2-c_s^2k^2})
\end{align}
We have shown in section \ref{sec:micro} that the above value of $c_s^2\leq1$ and thus, we see that according to the required conditions for causality in $(\omega,\vec{k})$ space, superluminal propagation of a DH photon is consistent with causality in the EFT regime.

\subsection*{Linear Path in Two Black Hole case}
For the superluminal photon propagating in the linear path in the two black hole case, we see that if our Green's function in position space is of the form $G_\text{R}(t,x^*)$, where $x^*$ is given in equation \eqref{eq:x*}, then the imaginary part of the retarded Green's function in $(\omega,\vec{k})$ space is given by the equation \eqref{imG}, as can be seen from the previous section. Thus, superluminality in this case too will be consistent with the microcausality in the EFT regime.

%% file: appendix.tex
\section{Calculation of commutators photon fields in curved spacetime}\label{appendix:A}
In this appendix, we will derive $A_\mu$ in the geometrical optics regime, which will lead us to explicitly find the commutator for both cases. We will follow the procedure given in \cite{Hui_2025} for this. The first thing that we will observe is that the equation of motion is given by:
\begin{align}
    \nabla_\mu F^{\mu\nu}-(2b\,R^{\mu}{}_\lambda  \nabla_\mu F^{\lambda\nu}+4c\,R^{\mu\nu\alpha\beta}\nabla_\mu F_{\alpha\beta})=0.
\end{align}
In the Lorenz gauge, we see that $\nabla_\mu A^\mu=0$. This leads to the equation of motion being:
\begin{align}\label{A.2}
    &\nabla^\mu\nabla_\mu A_\nu(1-2b\,R^{\lambda}{}_\lambda-8c\,R^{\mu\nu}{}_{\mu\nu})-2b(R^{\mu}{}_{\lambda}\nabla_\mu F^{\lambda}{}_{\nu}-R^{\lambda}{}_\lambda\nabla^\mu\nabla_\mu A_\nu)\notag\\
    -&8c(\,R^{\mu}{}_{\nu}{}^{\alpha\beta}\nabla_\mu F_{\alpha\beta}-\,R^{\mu\nu}{}_{\mu\nu}\nabla^\mu\nabla_\mu  A_{\nu})=0
\end{align}
 Now, using the geometric optics ansatz, we can substitute:
\begin{align*}
    A_\mu=Aa_\mu e^{i\Theta}
\end{align*}
and:
\begin{align*}
    k_\mu=\partial_\mu\Theta
\end{align*}
So that equation (\ref{A.2}) looks like:
\begin{align}\label{A.3}
    &a_\nu\nabla^\mu\nabla_\mu e^{i\Theta}(1-2b\,R^{\lambda}{}_\lambda-8c\,R^{\mu\nu}{}_{\mu\nu})-2b(R^{\mu}{}_{\lambda}\nabla_\mu f^{\lambda}{}_{\nu}-R^{\lambda}{}_\lambda a_\nu\nabla^\mu\nabla_\mu e^{i\Theta})\notag\\
    -&8c(\,R^{\mu}{}_{\nu}{}^{\alpha\beta}\nabla_\mu f_{\alpha\beta}-\,R^{\mu\nu}{}_{\mu\nu}a_\nu\nabla^\mu\nabla_\mu e^{i\Theta})=0
\end{align}
where, $f_{\mu\nu}=k_\mu a_\nu-k_\nu a_\mu$. Now, we can analyze each of the cases as given below
\subsection*{Circular path in Schwarzschild}
For the Schwarzschild metric, only the first term in equation (\ref{A.3}) survives. So, the equation of motion simplifies to:
\begin{align}\label{A.4}
    a_\nu\nabla^\mu\nabla_\mu e^{i\Theta}(1-8c\,R^{\mu\nu}{}_{\mu\nu})=0
\end{align}
Also note that the operator $\nabla^\mu\nabla_\mu$ acting on a scalar $s$ has the following property:
\begin{align}\label{A.5}
    \nabla^\mu\nabla_\mu s=\frac{1}{\sqrt{-g}}\partial_\mu(\sqrt{-g}g^{\mu\nu}\partial_\nu s)
\end{align}
Since $e^{i\Theta}$ is a scalar, we will use the above formula to simplify equation (\ref{A.4}).
\medskip

Now, for the circular path, we see that $k_r=k_\theta=0$ in Schwarzschild coordinates. Assuming the ansatz for $\Theta$ is given by:
\begin{align*}
    \Theta=\omega t-rk\phi
\end{align*}
This ansatz will lead to equation (\ref{A.4}) looking like:
\begin{align}
    a_\nu e^{i\Theta}((1-8c\,R^{t\nu}{}_{t\nu})\omega^2-(1-8c\,R^{\phi\nu}{}_{\phi\nu})k^2)=0
\end{align}
leading to the relation between $\omega$ and $k$ seen in equation \eqref{eq:Schwarzschild_omega}.
\medskip

Now, we would like to find the commutator $G_{\mu\nu}$. To do this, we will have to quantize the field $A_\mu$ as follows. We define $\hat A_\mu$ as:
\begin{align}
    \hat A_\mu=\int \frac{dk}{2\pi}\sum_{i=1}^2(a_\mu^{i,k}({x})e^{i\Theta}\hat{b}_{i,k}+{a_\mu^{i,k}({x})}^*e^{-i\Theta}\hat{b}_{i,k}^\dagger)
\end{align}
where $\hat{b}_{i,k}$ and $\hat{b}_{i,k}^\dagger$ are the annihilation and creation operators for a photon with polarization vector $a^i_\mu(k)$ and momentum $k$ in the circular path. Now, we require the following normalization condition on $\hat A_\mu$ so that the quantization condition for fields in \cite{Ashtekar:1975zn} is satisfied:
\begin{align}
    -ir^2F^{-1}(r)|a_\mu^{i,k}({x})|^2(e^{-i\omega t}\partial_te^{i\omega t}-e^{i\omega t}\partial_te^{-i\omega t})=1
\end{align}
for a non-zero $a_\mu^{i,k}({x})$. This leads to the value of $a^i_{\mu}$ when it is non-zero:
\begin{align}
    a_\mu^{i,k}({x})=\frac{F^{1/2}(r)}{r\sqrt{2\omega}}e^{i\alpha}
\end{align}
 where $\alpha$ is a phase factor. Now, if we require that equal time commutation relations hold, we automatically get the condition that:
 \begin{align}
     [\hat{b}_{i,k},\hat{b}_{j,k'}^\dagger]=\delta_{ij}\delta(k-k')
 \end{align}
 This leads to the commutator $G_{\mu\nu}$, being:
 \begin{align}
     G_{\mu\nu}(x,x')=[A_\mu({x}),A_\nu({x}')]=2i\sum_{i=1}^2\int a^{i,k}_{\mu}a^{i,k}_\nu \ \hbox{Im}(e^{i\omega(t-t')})e^{-ikr(\phi-\phi')}dk
 \end{align}
This is the form of the propagator in the EFT regime when geometric optics is valid. So taking a spatial Fourier transform, we get that $\tilde{G}_{\mu\nu}(k;t,t')$ is given by
\begin{align}
    \tilde{G}_{\mu\nu}(k;t,t')\sim i{F(r)}\frac{\sin(\omega (t-t'))}{\omega}\label{SchwarzchildG}
\end{align}
This gives us the structure of the commutator in the $(k,t)$ space, which we need for the microcausality bounds.

\subsection*{Linear path in the Two-Black-Hole case}
For this case too, we will see that on the linear path from $(x,0,0)$ to $(x',0,0)$, the second term in equation \eqref{A.3} is also 0. This means that, similar to the case of the circular path in the Schwarzschild metric, the equation of motion is again given by \eqref{A.4}. Using equation \eqref{A.5}, we see that the equation of motion looks like:
\begin{align}
    a_\nu [U^{2}(x)(1-8c\,R^{t\nu}{}_{t\nu}-2b\,R^{t}{}_t)\partial^2_te^{i\Theta}-U^{-2}(x)(1-8c\,R^{x\nu}{}_{x\nu}-2b\,R^{x}{}_x)\partial^2_xe^{i\Theta}]=0
\end{align}
Considering an ansatz of the form:
\begin{align*}
    e^{i\Theta}=e^{i\omega t}f(x)
\end{align*}
Substituting this into the equation, we get that
\begin{align}
    (1-8c\,R^{t\nu}{}_{t\nu}-2b\,R^{t}{}_t)f(x)\omega^2-U^{-4}(x)(1-8c\,R^{x\nu}{}_{x\nu}-2b\,R^{x}{}_x)f''(x)=0
\end{align}
Now, using the WKB approximation \cite{griffiths_introduction_2018}, we find that:
\begin{align}
    f(x)\approx e^{ig(x)}
\end{align}
where,
\begin{align}
    g(x)=\omega\int^xdx'\frac{1-4c\,R^{t\nu}{}_{t\nu}-b\,R^{t}{}_{t}}{1-4c\,R^{x\nu}{}_{x\nu}-b\,R^{x}{}_{x}}U^{2}(x')
\end{align}
Similar to how we quantized the photon field in the Schwarzschild background, we can quantize this field. This leads to us getting that
\begin{align}
    a^{i,k}_\mu(\vec{x})=U^{2}(x)\frac{1}{\sqrt{2k}}e^{i\alpha}
\end{align}
This means that the commutator $G_{\mu\nu}(x,x')$ is given by:
\begin{align}
    G_{\mu\nu}(x,x')=[A_\mu({x}),A_\nu({x}')]=2i\sum_{i=1}^2\int a^{i,k}_{\mu}a^{i,k}_\nu \ \hbox{Im}(e^{i\omega(t-t')})e^{-ik(x^*-{x'}^*)}dk
\end{align}
And we thus, get that:
\begin{align}
    \tilde{G}_{\mu\nu}(k;t,t')\sim i{U^{-2}(x^*)U^{-2}({x'}^*)}\frac{\sin(k (t-t'))}{k}
\end{align}
where, $U^{-2}(x^*)$ and $U^{-2}({x'}^*)$ are constants that only depend on the starting and ending position vector of the photon observed, and since both of these are less than 1, we see that 
\begin{align}
    |\tilde{G}_{\mu\nu}(k;t,t')|\leq\bigg|\frac{\sin(k (t-t'))}{k}\bigg|
\end{align}
which is the required form of the commutator $\tilde{G}_{\mu\nu}(k;t,t')$.